\documentstyle[12pt,psfig]{article}
\def\fsl#1{\setbox0=\hbox{$#1$}           % set a box for #1
   \dimen0=\wd0                                 % and get its size
   \setbox1=\hbox{/} \dimen1=\wd1               % get size of /
   \ifdim\dimen0>\dimen1                        % #1 is bigger
      \rlap{\hbox to \dimen0{\hfil/\hfil}}      % so center / in box
      #1                                        % and print #1
   \else                                        % / is bigger
      \rlap{\hbox to \dimen1{\hfil$#1$\hfil}}   % so center #1
      /                                         % and print /
   \fi}                                         %
\makeatletter
\def\@aabuffer{}
\def\author #1{\expandafter\def\expandafter\@aabuffer\expandafter
{\@aabuffer \small\rm      #1\relax \par}}
\def\address#1{\expandafter\def\expandafter\@aabuffer\expandafter
{\@aabuffer \small\it #1\relax \par\vspace{1em}}}

\def\maketitle{\newpage
\def\thefootnote{\fnsymbol{footnote}}
 \null
 {\normalsize \tt \begin{flushright}
  \begin{tabular}[t]{l} \@date
  \end{tabular}
 \end{flushright}}
\begin{center}
   {\bf \@title \par}
   \vskip 2em                      % Vertical space after title.
   \@aabuffer\relax
\end{center} \par
\gdef\@aabuffer{}
\setcounter{footnote}{0}
\def\thefootnote{\alph{footnote}}
}

\long\def\abstracts#1{
\begin{center}
{\begin{minipage}{4.2truein}
                 \footnotesize
                 \parindent=0pt #1\par
                 \end{minipage}}\end{center}
                 \vskip 2em \par}

\fussy
\def\section{\@startsection {section}{1}{\z@}{-3.5ex plus -1ex minus
    -.2ex}{2.3ex plus .2ex}{\bf }}
\def\subsection{\@startsection{subsection}{2}{\z@}
{-3.25ex plus -1ex minus
   -.2ex}{1.5ex plus .2ex}{\it }}

\def\thefootnote{\alph{footnote}}
\def\@makefnmark{{${}^{\@thefnmark}$}}

\renewenvironment{thebibliography}[1]
        {\begin{list}{\arabic{enumi}.}
        {\usecounter{enumi}\setlength{\parsep}{0pt}
         \setlength{\itemsep}{0pt}
         \settowidth
        {\labelwidth}{#1.}\sloppy}}{\end{list}}

\newcounter{arabiclistc}

\def\@citex[#1]#2{\if@filesw\immediate\write\@auxout
        {\string\citation{#2}}\fi
\def\@citea{}\@cite{\@for\@citeb:=#2\do
        {\@citea\def\@citea{,}\@ifundefined
        {b@\@citeb}{{\bf ?}\@warning
        {Citation `\@citeb' on page \thepage \space undefined}}
        {\csname b@\@citeb\endcsname}}}{#1}}

\newif\if@cghi
\def\cite{\@cghitrue\@ifnextchar [{\@tempswatrue
        \@citex}{\@tempswafalse\@citex[]}}
\def\citelow{\@cghifalse\@ifnextchar [{\@tempswatrue
        \@citex}{\@tempswafalse\@citex[]}}
\def\@cite#1#2{{$^{#1}$\if@tempswa\typeout
        {IJCGA warning: optional citation argument
        ignored: `#2'} \fi}}

\def\baselinestretch{0.975}
\ifx\selectfont\undefined
\@normalsize\else\let\glb@currsize=\relax\selectfont
\fi

\ifx\selectfont\undefined
\def\@singlespacing{%
\def\baselinestretch{1}\ifx\@currsize\normalsize\@normalsize
\else\@currsize\fi%
}
\else
\def\@singlespacing{\def\baselinestretch{1}\let\glb@currsize=
\relax\selectfont}
\fi

\long\def\@makecaption#1#2{
   \vskip 10pt
   \setbox\@tempboxa\hbox{\footnotesize #1: #2}
   \ifdim \wd\@tempboxa >\hsize   % IF longer than one line:
       \footnotesize #1: #2\par   %   THEN set as ordinary paragraph.
     \else                        %   ELSE  center.
       \hbox to\hsize{\hfil\box\@tempboxa\hfil}
   \fi}

\newcommand{\lsim}{\mbox{ \raisebox{-1.0ex}{$\stackrel{\textstyle <}
{\textstyle \sim}$ }}}
\makeatother
\begin{document}
\date{
  KEK TH-469\\
  KEK preprint 95-205\\
  February 1996}
%\date{}

\title{
  HIGGS PHYSICS AT $e^+e^-$ LINEAR COLLIDERS
\footnote{
    Talk presented
    at the Workshop on Physics and
    Experiments with Linear Colliders, Sep. 8--12, 1995,
    Morioka--Appi, Japan.
}
}
\author{
  YASUHIRO OKADA
\footnote{E-mail: {\tt okaday@theory.kek.jp}}
}
\address{
  National Laboratory for High Energy Physics (KEK) \\
  Oho 1-1, Tsukuba 305, Japan}
\maketitle\abstracts{
  Prospect of Higgs physics at future $e^+e^-$ linear colliders
  is reviewed for the weakly-coupled Higgs sector.
  Several topics related to the determination
  of various couplings of the Higgs boson in the standard model
  as well as in the supersymmetric standard model
  are discussed.
}

\section{Introduction}
It has been more than twenty years since the basic idea of the
standard model(SM) of elementary particle theory was proposed.
This theory is based on two fundamental principles,
{\em i.e.}\/ the gauge principle and the Higgs
mechanism.  From recent experiments it is now clear that strong and
electroweak interactions are described by an $SU(3) \times SU(2)
\times U(1)$ gauge theory.  On the other hand, little is known about
mechanism of the electroweak symmetry breaking.  To clarify dynamics
behind this symmetry breaking in the SM is the primary remaining
objective and the most important physics motivation of future
experiments at both LHC and $e^+e^-$ linear colliders.

Exploring the Higgs sector is important not only because the Higgs
particle is the only ingredient still missing from the SM, but
also because this will be a key to the physics beyond the SM.
In the minimal SM the Higgs sector consists of one Higgs doublet,
and the only free parameter is the Higgs-boson mass.  A heavy
Higgs boson corresponds to a large self-coupling
constant, and a light Higgs boson suggests that the dynamics
of the Higgs sector is well described perturbatively.  For many
extensions of the Higgs sector a similar relationship holds between the
Higgs-boson mass and the strength of the interaction which governs the
symmetry-breaking dynamics.  For example, if we require perturbative
unification of the
three gauge coupling constants as predicted in Grand Unified
Theories (GUT), the Higgs particle cannot be heavier than
about 200 GeV.  On the other hand, models like Technicolor predict
either a very heavy scalar particle or no
particle at all which acts as the Higgs boson.

In this talk I will discuss the physics of the Higgs sector at
future $e^+e^-$ linear-collider experiments with a center-of-mass (CM)
energy ranging from 300 GeV to 1.5 TeV.  I restrict myself to
the ``light Higgs case'' where at least one Higgs particle
exists below, say, 200 GeV, and the dynamics
behind this particle is described by perturbation theory.
The strategy to
study the strongly coupled Higgs sector in future $e^+e^-$
linear-collider
experiments is quite different from that in the light Higgs
case, and it is covered elsewhere in this workshop.\cite{SCHiggs}
{}From intensive discussions, including those reported in the
last workshop of this series,\cite{Gunion,Janot,Kawagoe}
it is now clear that the discovery of such a Higgs particle
is easy at an  $e^+e^-$ linear collider
with $\sqrt{s}\sim 300 - 500$ GeV,
if it exists with a mass below 200 GeV
and with a production cross-section and decay
branching ratios similar to the SM Higgs boson.
Therefore I would like to stress subjects which become
important after the Higgs particle is discovered.
In other words, I would like to consider how various
couplings related to the
Higgs particle will be measured in future $e^+e^-$ experiments
and what the impact of these measurements will be on the
establishment of the SM and
the search for physics beyond the SM.  Many of the contributions
in the Higgs session in this workshop are related to various Higgs
couplings.

In this talk I will first give a short review of the SM Higgs sector
and the Higgs sector in the Minimal Supersymmetric
Standard Model (MSSM). Also I will comment on Higgs mass and
properties for some extended versions
of supersymmetric (SUSY) standard models.
Then, I will report four topics
discussed at this workshop, all of which are related to measurements of
various Higgs couplings.

\section{The SM Higgs}
In the minimal SM the Higgs-boson mass ($m_h$) and the Higgs-boson
self-coupling constant ($\lambda$) from the Higgs potential
($V_{Higgs}=m^2 |\Phi|^2 + \lambda|\Phi|^4$,
$(m^2\le0)$) are related by $m_h^2 = 2\lambda \upsilon^2$,
where $\upsilon (= 246$ GeV) is the vacuum expectation value.
As a result, the theory
behaves quite differently for small and large values of the
Higgs-boson mass.
This situation can be most easily understood by evaluating the running
self-coupling constant, $\lambda$, using the renormalization group
equations (RGE's).
Neglecting Yukawa coupling constants except for the one for
the top quark,
$y_t$, the RGE for $\lambda$ at the one-loop level is written as
\begin{equation}
\frac{d\lambda}{dt} = \frac{1}{(4\pi)^2}
\{24\lambda^2 + 12y_t^2\lambda - 6y_t^4-12A\lambda+6B\},
\end{equation}
where $A = \frac{1}{4}g_1^2 + \frac{3}{4}g_2^2$,
$B = \frac{1}{16}g_1^4 + \frac{1}{8}g_1^2g_2^2 + \frac{3}{16}g_2^4$
for $U(1)$ and $SU(2)$ gauge coupling constants $g_1$ and $g_2$
respectively and $t=\ln{\mu}$ ($\mu$ is the renormalization scale).
Since the input values of the gauge coupling
constants and the top Yukawa
coupling constant at the electroweak scale can be determined from
recent experimental results, we can draw the flow of the running
self-coupling constant for each value of $m_h$ as shown in Figure 1.
%Fig1
%
\begin{figure}
\begin{center}
\mbox{\psfig{figure=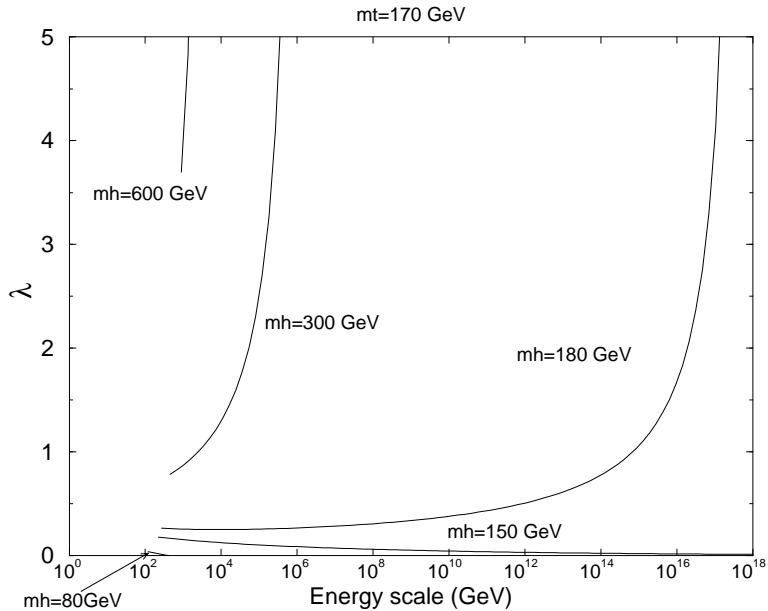,width=4in,angle=-90}}
\end{center}
\caption{The flow of the Higgs self-coupling constant ($\lambda$)
in the minimal SM for the several values of the Higgs
boson masses. The top quark mass is assumed to be 170 GeV.
\label{fig:fig1}}
\end{figure}

 From this figure it is clear that the possible scenarios for new
 physics are different for
 two cases, {\em i.e.}\/ (i) $m_h \gg m_{top}$ and
(ii) $m_h \lsim m_{top}$.
 In case (i) the coupling constant becomes very large at a relatively
 low energy scale, and therefore new physics is indicated well
 below the Planck scale ($\simeq 10^{19}$ GeV).  On the other hand,
 the theory can be
 weakly coupled up to approximately the Planck scale in case (ii),
 which is consistent
 with the idea of grand unification.  Of course the flow
of the coupling
constants is different if we change the particle content
in the low-energy
theory, but the upper bound on the Higgs mass is believed
to be about 200 GeV for most GUT models.
This is also true for the SUSY GUT.
As for the MSSM, however, a stronger bound on the Higgs mass
is obtained
independently of the assumption of grand unification.

\section{The Higgs Sector in the MSSM}
Supersymmetry is now considered as the most promising candidate
for physics beyond the SM.  Recent improvements in measurements
of the three gauge coupling
constants enable us to distinguish various  GUT models, and
it has become
clear that the supersymmetric version is favored.\cite{unification}
The particle content of the SUSY GUT below the GUT scale is
just that of the MSSM.

The Higgs sector of the MSSM consists of
two Higgs doublets.\cite{review}
The most important feature of this Higgs sector is
that the Higgs-self-coupling
constant at the tree level is
completely determined by the $SU(2)$ and $U(1)$ gauge coupling
constants. After electroweak symmetry breaking, the physical
Higgs states include two CP-even Higgs bosons ($h, H$),
one CP-odd Higgs boson ($A$) and one pair of charged Higgs
bosons ($H^\pm$) where we denote by $h$
and $H$ the lighter and heavier Higgs bosons respectively.  Although
at the tree level the
upper bound on the lightest CP-even Higgs boson mass is given
by the $Z^0$ mass, the radiative corrections
weaken this bound.\cite{OYY} The Higgs potential is given by

\begin{eqnarray}
V_{Higgs} & = & m^2_1|H_1|^2 + m_2^2|H_2|^2 - m_3^2
   (H_{1}\cdot H_{2}+\bar{H_{1}}\cdot\bar{ H_{2}})
\nonumber\\
          & & +\frac{g_2^2}{8}(\bar{H_1}\tau^aH_1
                + \bar{H_2}\tau^aH_2)^2
                + \frac{g_1^2}{8}(|H_1|^2 - |H_2|^2)^2
\nonumber\\
          & & + \Delta V,
\end{eqnarray}
where $\Delta V$ represents the contribution from one-loop diagrams.
Since the loop correction due to the top quark and its
superpartner, the stop squark, are proportional to the fourth
power of the top Yukawa coupling constant and hence are large, the
Higgs self-coupling constant is no longer determined only by
the gauge coupling constants.
The upper bound on the lightest CP-even Higgs mass ($m_h$)
can significantly increase for a reasonable choice of the
top-quark and stop-squark masses. Figure 2 shows
the upper bound on $m_h$ as a function of top-quark mass
for several choices of the stop mass and the ratio of two Higgs-boson
vacuum expectation values ($\tan\beta = \frac{<H_2^0>}{<H_1^0>})$.
%Fig2 (uppernound on the lightest Higgs mass in MSSM)
%
\begin{figure}
\begin{center}
\mbox{\psfig{figure=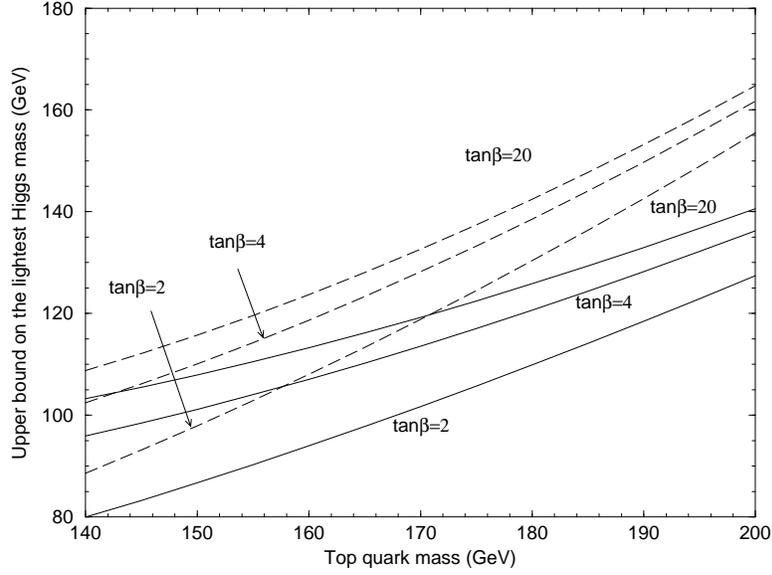,width=4in,angle=-90}}
\end{center}
\caption{The upper bound on the lightest CP-even Higgs mass in the MSSM
as a function of the top quark mass for various $\tan{\beta}$ and
two large stop mass scales. The solid (dashed) line corresponds
to $m_{stop}$=1 (10) TeV without left-right mixing of two stop states.
These masses are calculated by the method with the renormalization
group equation.\protect\cite{OYYRGE}
\label{fig:fig2}}
\end{figure}
We can see that, in the MSSM, at least one neutral Higgs-boson should
exist below 130 - 150 GeV depending on the top and stop masses.

Other Higgs states, namely the $H, A, H^\pm$, are also important
to clarify the structure of the model. Their existence alone is proof
of new physics beyond the SM, but we may be able to distinguish the
MSSM from a general two-Higgs model through the investigation of
their masses and couplings.  In the MSSM the Higgs sector is
described by four independent parameters for which we take the
mass of the CP-odd Higgs boson ($m_A$), $\tan\beta,$ the top-quark mass
$(m_t$) and the stop mass ($m_{stop}$).
The top and stop masses enter through radiative corrections to
the Higgs potential.  Speaking precisely, there are left- and
right-handed stop states which can mix to form
two mass eigenstates; therefore more than just one parameter
is required to specify the stop sector.
In Figure 3, the masses for the $H, A$, and $ H^\pm$ are shown as a
function of $m_A$ for several choices of $\tan\beta$
and $m_{stop}$=1 TeV.
%Fig 3 (Higgs masses as a function of ma)
%
\begin{figure}
\begin{center}
\mbox{\psfig{figure=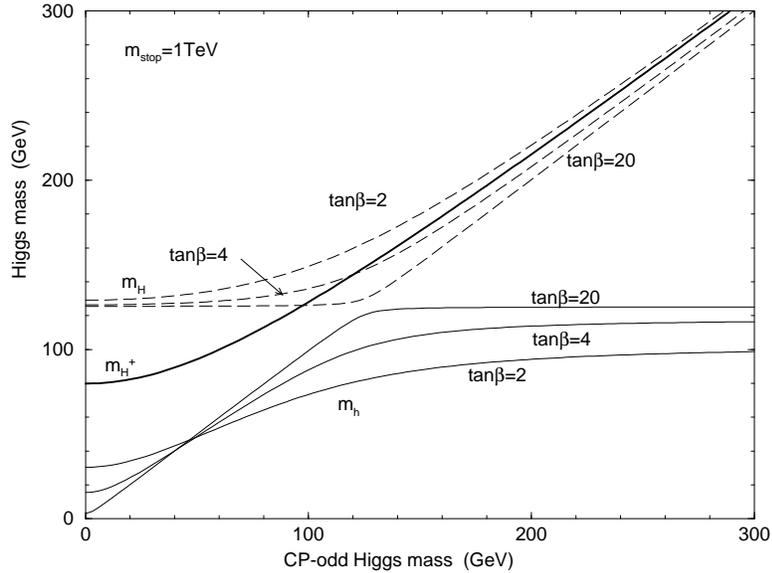,width=4in,angle=-90}}
\end{center}
\caption{The light ($h$), heavy ($H$) CP-even Higgs masses and
the charged
Higgs ($H^\pm$) mass as a function of the CP-odd Higgs ($A$) mass.
The top and stop masses are taken as $m_t$ =~170 GeV and
$m_{stop}$ =~1 TeV.
\label{fig:fig3}}
\end{figure}
We can see that, in the limit of $m_A \rightarrow \infty$, $m_h$
approaches a constant value which corresponds to the upper bound
in Figure 2. Also in this limit the
$H, A$ and $H^\pm$ become degenerate in mass.

The neutral Higgs-boson couplings to gauge bosons and fermions
are determined by the ratio of vacuum expectation values $\tan{\beta}$
and the mixing angle $\alpha$
of the two CP-even Higgs particles defined as
\begin{eqnarray}
ReH_1^0 &=& \frac{1}{\sqrt{2}}
(\upsilon\cos\beta - h\sin\alpha + H\cos\alpha)
\nonumber \\
ReH_2^0 &=& \frac{1}{\sqrt{2}}
(\upsilon\sin\beta + h\cos\alpha + H\sin\alpha).
\end{eqnarray}
For Higgs-boson production, the Higgs-bremsstrahlung process
$e^+e^- \rightarrow Zh$ or $ZH$ and the associated production
$e^+e^- \rightarrow Ah$ or $AH$ play complimentary roles.
Namely $e^+e^- \rightarrow Zh~(ZH)$ is proportional
to $\cos(\beta-\alpha)(\sin(\beta-\alpha))$, and
$e^+e^- \rightarrow Ah~(AH)$ is proportional to
$\sin(\beta-\alpha)(\cos(\beta-\alpha))$,
so at least one of the two processes has a sizable coupling.
It is useful to distinguish the following two cases when we discuss
the properties of the Higgs particles in the MSSM, namely
(i) $m_A \lsim 150$ GeV, (ii) $m_A \gg 150$ GeV.
In case (i), the two CP-even Higgs bosons can have large mixing, and
therefore the properties of
the neutral Higgs boson can be substantially different from those
of the minimal SM Higgs.  On the other hand, in case (ii), the
lightest CP-even Higgs becomes a SM-like Higgs, and the other
four states, $H, A, H^\pm$ behave as a Higgs doublet orthogonal
to the SM-like Higgs doublet.  In this region,
$\cos(\beta-\alpha)$ approaches unity and $\sin(\beta-\alpha)$
goes to zero so that $e^+e^- \rightarrow Zh$ and
$e^+e^- \rightarrow AH$ are the dominant production processes.
Scenarios for the Higgs physics at a future $e^+e^-$ linear
collider are different for two cases.  In case (i) it is possible
to discover all Higgs states with $\sqrt{s} = 500$ GeV, and the
production cross-section of the lightest Higgs boson may be
quite different from that of the SM so that
it may be clear that the discovered Higgs is not the SM Higgs.
On the other hand, in case (ii), only the lightest Higgs may
be discovered at the earlier stage of the $e^+e^-$ experiment,
and we have to go to a higher energy machine to find the heavier
Higgs bosons. Also, since the properties
of the lightest Higgs boson may be quite similar to those of
the SM Higgs boson we need precision experiments on the production
and decay of the particle
in order to investigate possible deviations from the SM.

\section{The Higgs-boson mass and production cross-section in
extended versions of the SUSY SM}
Although the MSSM is the most widely studied model, there are
several extensions of the SUSY version of the SM.  If we focus
on the structure of the Higgs sector, the MSSM is special because
the Higgs self-couplings at the tree level
are completely determined by the gauge coupling constants.
It is therefore important to know how the Higgs phenomenology
is different for models other than the MSSM.

A model with a gauge-singlet Higgs boson is the simplest
extension.\cite{singlet}  This
model does not destroy the unification of the three gauge coupling
constants since the new light particles do not carry the SM quantum
numbers. Moreover, we can include a term $W_\lambda = \lambda NH_1H_2$
in the superpotential where $N$ is a gauge singlet superfield.
Since this term induces $\lambda^2|H_1H_2|^2$ in the Higgs potential,
the tree-level Higgs-boson self-coupling depends on $\lambda$
as well as the gauge coupling constants.  There is no definite
upper-bound on the lightest CP-even Higgs-boson mass in this model
unless a further assumption on the strength of the coupling $\lambda$
is made.  If we require all dimensionless
coupling constants to remain perturbative up to the GUT scale
we can calculate the upper-bound of the lightest CP-even Higgs-boson
mass.\cite{singletmass} In Figure 4, the upper bound of
the Higgs-boson mass is shown as a function of the top-quark mass.
%Fig4. (NMSSM Higgs mass)
%
\begin{figure}
\begin{center}
\mbox{\psfig{figure=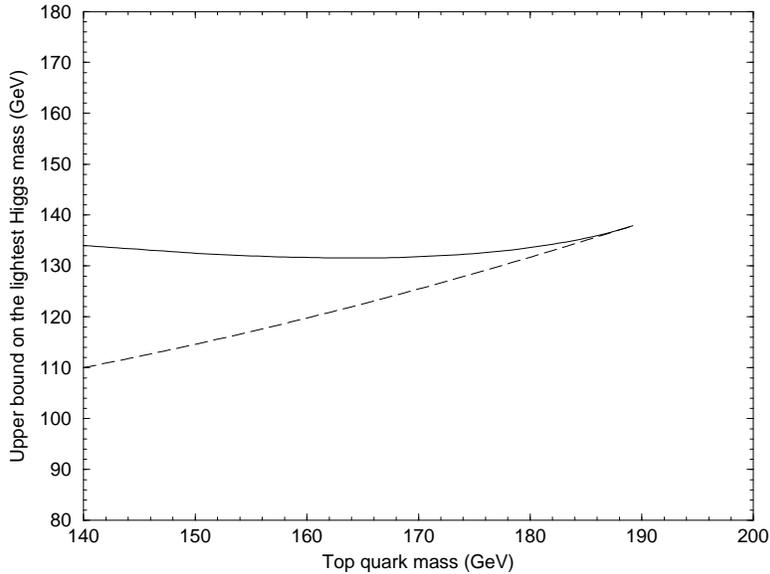,width=4in,angle=-90}}
\end{center}
\caption{The upper bound on the lightest CP-even Higgs mass
in the SUSY SM with a gauge singlet Higgs (the solid line).
The stop mass is taken as 1 TeV. The dotted line corresponds
to the upper bound in the MSSM case.
\label{fig:fig4}}
\end{figure}
In this figure we have taken the stop mass as 1 TeV and demanded
that no dimensionless coupling constant may blow up below the GUT scale
($\sim 10^{16}$ GeV).  We can see that the upper bound is given
by 130 $\sim$ 140 GeV for this choice of the stop mass.
The top-quark-mass dependence is not significant compared to
the MSSM case because the
maximally allowed value of $\lambda$ is larger (smaller) for a
smaller (larger) top mass.

 From this figure we can see that the lightest Higgs boson is at least
kinematically accessible at an $e^+e^-$ linear
collider with $\sqrt{s}\sim 300 - 500$ GeV.  This does not, however,
mean that the lightest Higgs boson is detectable.  In this model
the lightest Higgs boson is composed of one gauge singlet and two
doublets, and if it is singlet-dominated its couplings to the
gauge bosons are significantly reduced, hence its
production cross-section is too small. In such a case the heavier
neutral Higgs bosons may be detectable since these bosons have a
large enough coupling to gauge bosons. In fact we can put an
upper-bound on the mass of the heavier Higgs boson when the
lightest one becomes singlet-dominated.
By quantitative study of the masses and the production
cross-section of the Higgs bosons in this model, we can show that
at least one of the three CP-even Higgs bosons has a
large enough production cross-section in the
$e^+e^- \rightarrow Zh^o_i$ $(i =1, 2, 3)$ process to be detected
at an $e^+e^-$ linear collider with
$\sqrt{s}\sim300 - 500$ GeV.\cite{KOT}  For this purpose we define
the minimal production
cross-section, $\sigma_{min}$, as a function of $\sqrt{s}$ such that
at least one of these three $h^0_i$  has a larger production
cross section than $\sigma_{min}$ irrespective of the
parameters in the Higgs mass matrix.
We can show that $\sigma_{min}$ is  larger than 0.04 pb for
$m_t = 120 -  180$ GeV and $m_{stop} = 1$ TeV at an $e^+e^-$
linear collider with $\sqrt{s}=300$ GeV, and therefore the
discovery of at least one neutral
Higgs boson is guaranteed with an integrated luminosity of 10 fb$^{-1}$.
More recently, a condition to give $\sigma_{min}$ is improved
by a closer investigation, and $\sigma_{min}$ turns
to be  given by just one third of the SM production cross-section
with the Higgs boson mass equal to the upper-bound value.\cite{King}
In Figure 5 we show this $\sigma_{min}$ as a function of $\sqrt{s}$.
%Fig5. (sigma min for NMSSM)
\begin{figure}
\begin{center}
\mbox{\psfig{figure=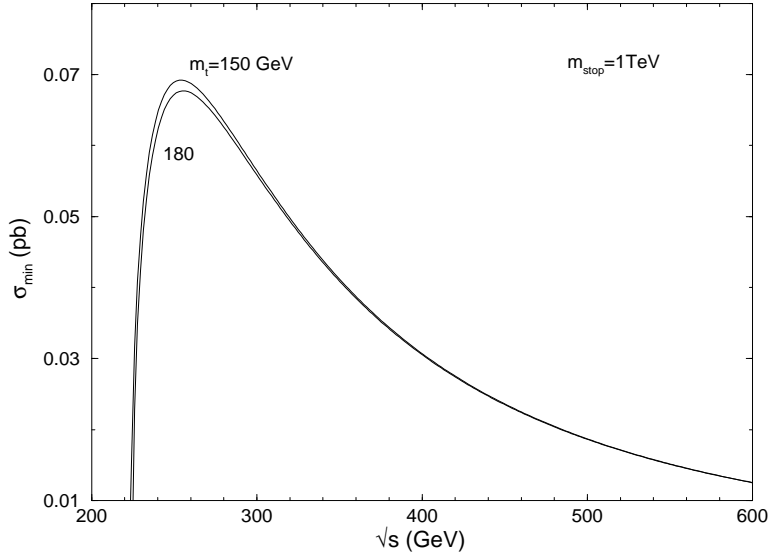,width=4in,angle=-90}}
\end{center}
\caption{Minimal production cross section, $\sigma_{min}$, for
the SUSY SM with a gauge singlet Higgs for the top mass
$m_t$=150 and 180 GeV and $m_{stop}$=1 TeV.
\label{fig:fig5}}
\end{figure}
If we include more singlets
$\sigma_{min}$ is just given by $\frac{1}{n+2}$ times the SM
production cross-section, where $n$ is the number of gauge-singlet
Higgs bosons which mix with the doublets.
Therefore, as long as the number of gauge-singlet Higgs bosons is not
too large ($\lsim$5), it is possible to discover at least one
neutral Higgs boson in the first stage of the  $e^+e^-$
linear-collider experiment.

\section{A ZZh anomalous coupling at a TeV linear collider}
In order to establish the SM and to search for new physics beyond
it, it is fundamentally important to investigate whether or not
various couplings among gauge bosons and the Higgs boson are
described by the SM. For example, the existence of the self-couplings
of gauge bosons are an important feature of the non-Abelian nature of
the gauge interaction.

If a Higgs boson is discovered at a relatively light mass scale
($\lsim 200$ GeV),
most probably the physics of electroweak
symmetry breaking is described by an interaction whose strength
is not very much different from that of the electroweak interaction.
In such a case the measurement of the couplings involving the light
Higgs boson is interesting in order to look for effects from
physics beyond the SM. Assuming that some new physics exists
in the multi-TeV region, we can write
down the general form of higher dimensional operators which are
induced after integrating out the heavy fields.
\begin{equation}
L = \sum_{i}\frac{f_i}{\Lambda^2}O_i +\ldots
\end{equation}
where the $f_i$ are dimensionless couplings, and $\Lambda$ is the
new-physics scale.
The $O_i$ are gauge-invariant operators composed of gauge bosons and
Higgs doublet fields as well as fermion fields.\cite{linear}

In this workshop it was pointed out that the production
cross-section in the Higgs-bremsstrahlung process,
$e^+e^- \rightarrow Zh$, is sensitive to
one anomalous coupling of this type, {\em i.e.}
$\frac{f}{\Lambda^2}(D_\mu\Phi)^\dagger W^{\mu\nu}(D_\nu\Phi)$,
for an $e^+e^-$ linear collider with a center-of-mass energy
over 1 TeV.\cite{Djouadi}
In the SM in this energy region the dominant Higgs production process
is WW fusion process
rather than Higgs-bremsstrahlung.
However, as long as the new-physics effect are concerned, the
Higgs-bremsstrahlung process is more important.
This is because the relevant energy scale of
this process is $\sqrt{s}$ and the anomalous coupling
of $ZZh$ becomes large as the energy scale increases while the
energy scale relevant to the fusion process is the Higgs mass scale,
not $\sqrt{s}$. Thus, in order to look for new-physics effects,
the measurement of the production cross-section
in $e^+e^- \rightarrow Zh$ is more important than the process
$e^+e^- \rightarrow WW\bar{\nu}\nu \rightarrow h\bar{\nu}\nu$.

\section{Determination of the heavy Higgs mass scale
from branching measurements in the MSSM}
It is generally accepted that the SM Higgs boson  with a mass
less than about 200 GeV will be discovered at the first stage
of an $e^+e^-$ linear collider
experiment where the CM energy is $\sim 300 -  500$ GeV.
This is sufficient to discover at least one CP-even Higgs boson
of the MSSM.  If a Higgs boson is discovered, we would like
to determine whether or not this boson is the SM Higgs boson
by studying its production and decays.
It is therefore important to investigate to what extent the production
cross-section and decay branching ratios can be determined and
what the impact of these determinations will be on establishing the
SM and searching for physics beyond the
SM.\cite{Janot,Kawagoe,JLC,Hildreth}
In the context of the MSSM, the question can be restated as whether
the parameters in the Higgs sector are determined by various
observable quantities related to the Higgs boson. Although it is
possible to discover all five Higgs
states at the first stage of the linear collider experiment,
we may at first be able to find only one CP-even Higgs boson.
In this situation it is important to determine in
which mass region the other Higgs states exist so that
these particles become targets of the second stage of
the $e^+e^-$ linear-collider experiments after the beam energy
is increased.

This problem was addressed by Kamoshita's talk in this
workshop.\cite{Kamoshita}
The free parameters required to specify the Higgs sector in the MSSM
can be taken
to be the CP-odd Higgs-boson mass ($m_A$), the ratio of
two vacuum expectation values ($\tan\beta$) and
masses of the top quark and the stop squark. The latter
two parameters ($m_t, m_{stop}$) are necessary to evaluate
the Higgs potential at the one-loop level.
Suppose that the lightest CP-even Higgs boson is discovered
such that its mass ($m_h$) is precisely known.
Then we can solve for one of the free
parameters, for example, $\tan\beta$, in terms of the
other parameters. Assuming the top-quark mass is well determined
by the time when the $e^+e^-$ linear
collider is under operation, the unknown parameters for
the Higgs sector are then $m_A$ and $ m_{stop}$.  The question is,
to what extent these parameters are
constrained from observable quantities such as the production
cross-section and the various branching ratios.

It has been pointed out that one particular ratio of two
branching ratios,
\begin{equation}
R_{br}\equiv \frac{Br(h \rightarrow c\bar{c}) + Br(h \rightarrow gg)}
{Br(h \rightarrow b\bar{b})},
\end{equation}
is especially useful to constrain the heavy Higgs mass scale.
In the MSSM, each of the two Higgs doublets couples to either
up-type or down-type quarks.  Therefore, the ratio of the Higgs
couplings to up-type quarks and to down-type quarks is
sensitive to the parameters of the Higgs sector, {\em i.e.} the angles
$\alpha$ and $\beta$ in Section 3.
Since the gluonic width of the Higgs boson is generated by a
one-loop diagram with an internal top-quark,
the Higgs-gluon-gluon coupling is essentially
proportional to the Higgs-top coupling.
Then $R_{br}$ is proportional to square of the ratio of
the up-type and down-type Yukawa coupling constants.  Since
the up-type (down-type) Yukawa coupling constant contains a
factor $\frac{\cos\alpha}{\sin\beta}$,
$(-\frac{\sin\alpha}{\cos\beta})$ compared to the SM coupling constant,
$R_{br}$ is proportional to $(\tan{\alpha} \tan{\beta})^{-2}$.
In Figure 6 $R_{br}$ is shown as a function of $m_A$ for several
choices of $m_{susy}(\equiv m_{stop})$.
%Fig.6 (R_br)
%
%
\begin{figure}
\begin{center}
\mbox{\psfig{figure=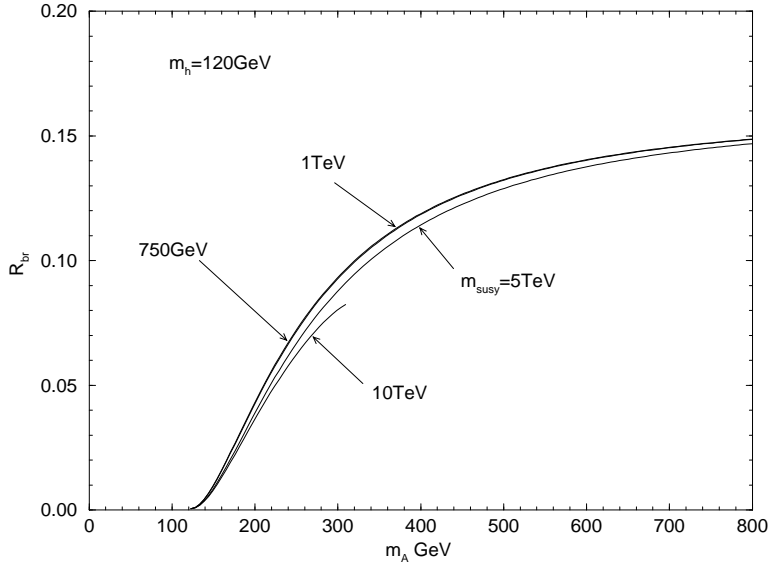,width=4in,angle=-90}}
\end{center}
\caption{
$R_{br}\equiv
\protect\frac{(Br(h\to c\bar{c})+Br(h\to gg))}{Br(h\to b\bar{b})}$
as a function of $m_A$ for several values of $m_{susy}$
for the lightest CP-even Higgs
mass $m_h=120$ GeV.\protect\cite{Kamoshita}
The following parameters are used for the calculation of
the branching ratios: $m_t= 170$ GeV, $\bar m_c(m_c)=1.2$ GeV,
$\bar m_b(m_b)=4.2$ GeV, $\alpha_s(m_Z)=0.12$.
\label{fig:fig6}}
\end{figure}
{}From this figure we can see that $R_{br}$ is almost independent of
$m_{stop}$.  In fact, it can be shown that $R_{br}$ in the MSSM,
normalized by $R_{br}$ in the SM, is
approximately given by,
\begin{equation}
\frac{R_{br}(MSSM)}{R_{br}(SM)}\approx \left(\frac{m_h^2 - m_A^2}
{m_Z^2 + m_A^2}\right)^2
\end{equation}
for $m_A \gg m_h \sim m_Z$.  Measuring this quantity to a good accuracy
is therefore important for constraining the scale of the heavy
Higgs mass. Note that $R_{br}$ approaches the SM value in the
large $m_A$ limit. We can see that $R_{br}$ is reduced by 20$\%$
even for $m_A = 400$ GeV.

In Nakamura's talk the experimental determination of these
branching ratios was discussed.\cite{Nakamura}
Although it is very difficult to measure the charm and gluonic
branching ratios separately with good accuracy, the sum of the
two branching ratios can be determined reasonably well. The
statistical error in the determination of $R_{br}$ after two
years at an $e^+e^-$ linear collider with $\sqrt{s} = 300$ GeV
is 19$\%$. We also need to know the theoretical ambiguity of the
calculation of the branching
ratios in $h \rightarrow b\bar{b}, c\bar{c}, gg$.
Uncertainties in the
charm-quark and bottom-quark masses as well as in
the strong coupling constant are important.
At the moment the theoretical error in the calculation of $R_{br}$
is estimated to be larger than 20$\%$ and mainly comes from
uncertainties in $\alpha_s$ and $m_c$.\cite{Kamoshita,DSZ}
Both theoretical and experimental improvements are necessary
to calculate the
branching ratios more precisely.

\section{Heavy Higgs decays to SUSY particles}
The investigation of the properties of the heavy
Higgs bosons ($A, H, H^{\pm}$) is one of main
goals of a TeV linear collider.  Since the MSSM is a special type
of two-Higgs-doublet model, the discovery of these particles
as predicted is strong evidence for the MSSM.
Also, the determination of the parameters
$m_A$ and $\tan\beta$ of the Higgs sector
is important in exploring the whole structure of the SUSY model since
these parameters are relevant to other sectors of the model
in the context of
the MSSM and/or supergravity models.
If we assume that the mass of the heavy Higgs is larger
than $200\sim300$ GeV
it is possible that some decay channels to SUSY particles are open.
In Djouadi's talk various decays of heavy Higgs bosons including
SUSY modes are considered in the context of the SUSY GUT model with Yukawa
unification.\cite{Djouadi}

Let us first summarize the dominant decay modes of heavy Higgs bosons
if SUSY decay modes are not open.
Since the coupling of the $H$ and the $A$ to down-type quarks is
enhanced for large $\tan\beta$, the $H$ and $A$ dominantly decay to
$b\bar{b}$ or $\tau^+\tau^-$ for $\tan\beta > 10$.
The situation is different in the smaller $\tan\beta$ region
where, if the $t\bar{t}$ mode is open, this mode
dominate over other modes. Below the $t\bar{t}$ threshold
the $H \rightarrow hh, H \rightarrow WW$
and $A \rightarrow Zh$ modes can be dominant.  For the charged
Higgs boson, the
$H^+ \rightarrow \bar{b}t$ mode is dominant if kinematically
accessible, and otherwise
$H^+ \rightarrow \tau^+\nu$ becomes the main decay mode.

If we allow SUSY decay modes, heavy Higgs bosons can decay
to squark-pairs,
slepton-pairs and charginos and neutralinos. Of course whether or not
these decay modes are available depends on the mass spectrum of SUSY
particles.  Here a model
based on minimal supergravity is considered with an assumption
of Yukawa coupling unification.\cite{Yukawa}
Requiring that the $m_b/m_\tau$ ratio is correctly reproduced from
the SU(5) SUSY GUT assumption and that
the electroweak symmetry breaking is induced from
the renormalization effects on the Higgs mass term from
the universal SUSY-breaking mass at the GUT scale
(the radiative electroweak symmetry breaking
scenario),\cite{radiative} we can reduce the number
of free parameters of the model.  There are two separate
regions of $\tan\beta$ according to this scenario, but the
so-called small $\tan\beta$ solution is
interesting where $\tan\beta \simeq 1.75$.  Essentially
this model contains two free parameters for which we can
take $m_A$ and $M_{1/2}$ (a gaugino mass parameter).  Then,
for a fixed heavy Higgs mass, all superparticle masses
are determined as a function $M_{1/2}$ so that the decay widths
including SUSY modes are calculable.  The importance of the
SUSY modes is quite different depending upon whether or not
the $H(A) \rightarrow t\bar{t}$ mode is open.
Below the top threshold the SUSY modes can dominate over
the SM mode, and the total width can be enhanced by an order
of magnitude.  This is especially evident if the stop becomes
light enough to be a dominant decay mode.  On the
other hand, above the top threshold the $t\bar{t}$ mode is almost
always dominant, and the SUSY modes play minor roles.  This
is because the decay width to the $t\bar{t}$ pair is
already large compared to other
modes.  Only when the stop is light enough can the SUSY modes
give a sizable contribution to the total width.

\section{Multi-Higgs production in the MSSM}
There were two talks which covered multi-Higgs production
in the MSSM; one deals with various double and triple
Higgs production processes at a TeV collider\cite{Djouadi,DHZ},
and the other focuses on the $e^+e^- \rightarrow Zhh$ process at
an $e^+e^-$ linear collider
with $\sqrt{s}\sim 300 - 500$ GeV.\cite{Watanabe}

In the SM the $e^+e^- \rightarrow Zhh$ process is especially
important because this process depends on the triple Higgs
coupling (Figure 7).\cite{SMZHH}
%Fig.7
%
\begin{figure}
\begin{center}
\mbox{\psfig{figure=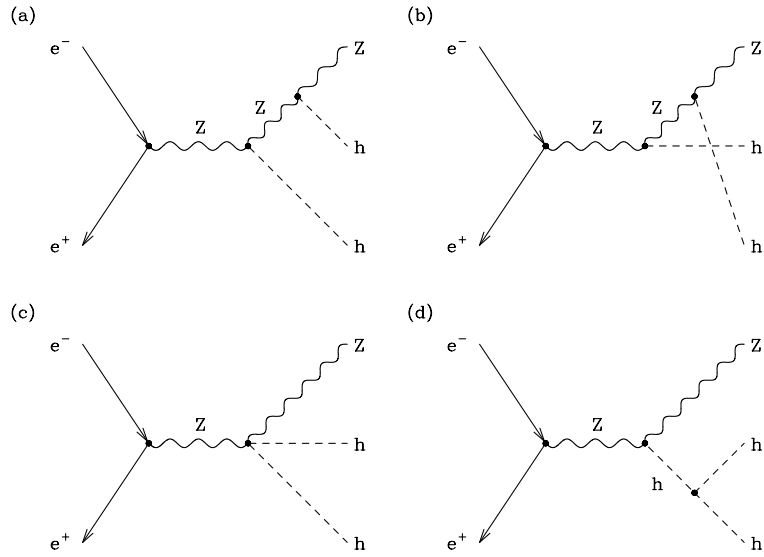,width=4.5in}}
\end{center}
\caption{
  The relevant Feynman diagrams for $e^+e^-$ $\rightarrow Zhh$
in the SM.
\label{fig:fig7}}
\end{figure}
Therefore, information on
the Higgs potential is obtained from this process.  The production
cross-section is, however, not so large.  For $m_h = 100$ GeV
it is a few times
$10^{-1}$ fb for $\sqrt{s} = 500 \sim 1$ TeV; therefore,
we need more than a hundred fb$^{-1}$ to observe this process.
In the MSSM case the situation changes in
two ways.  The $ZZh$ and the $h^3$ couplings are modified from the SM
couplings, and additionally there are diagrams which contain
heavy Higgs($H,A$) in an internal line as shown in Figure 8.
%Fig.8
%
\begin{figure}
\begin{center}
\mbox{\psfig{figure=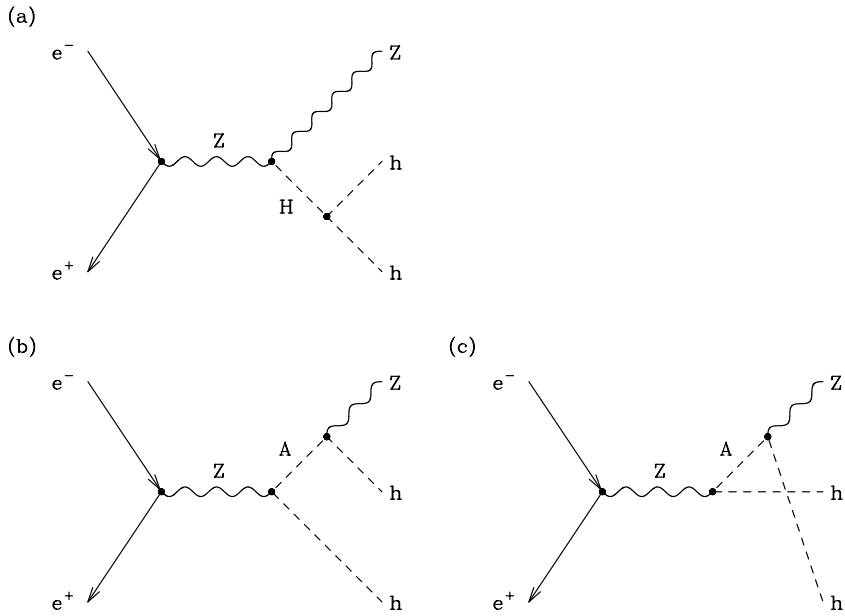,width=5in}}
\end{center}
\caption{ The additional Feynman diagrams for $e^+e^-$
$\rightarrow Zhh$ in the MSSM.
\label{fig:fig8}}
\end{figure}
In both talks it was noticed that the multi-Higgs production
cross-section becomes large only when both heavy Higgs
($H$, $A$) production and the
subsequent decay through $H \rightarrow hh$ or $A \rightarrow hZ$
are kinematically allowed.
In Figure 9, the production cross-section
$\sigma(e^+e^- \rightarrow Zhh)$
is given as a function of $m_A$ for $\tan\beta$ = 2 and 10
for the MSSM.
%Fig.9
%
\begin{figure}
\begin{center}
\mbox{\psfig{figure=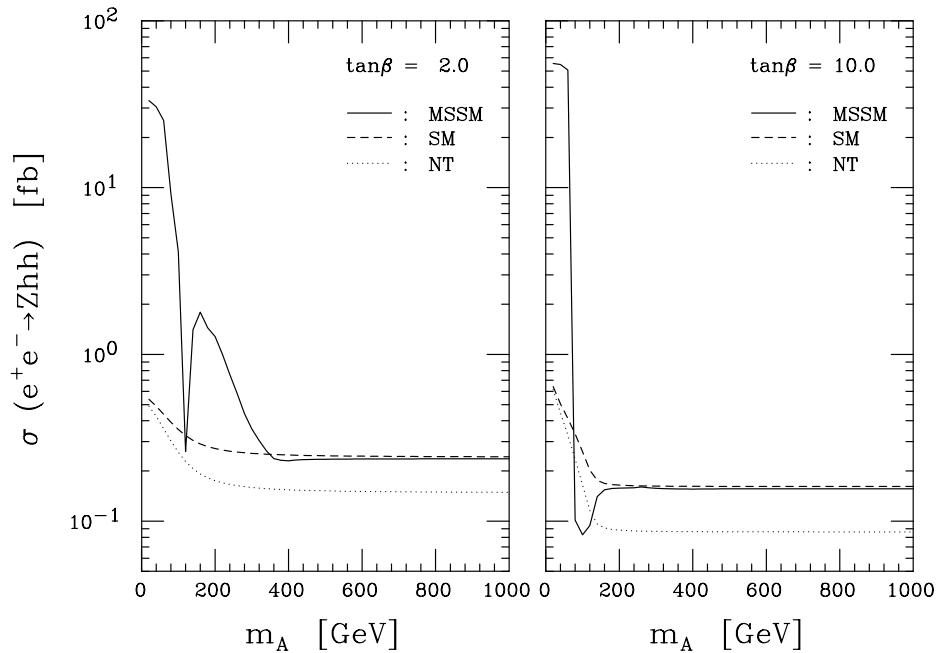,width=5in}}
\end{center}
\caption{
The $m_A$ dependence of the $e^+e^-$ $\rightarrow Zhh$
cross-section in the MSSM (solid) for $\tan \beta$ =~2 and 10
at $\protect\sqrt{s}$ = 500 GeV.\protect\cite{Watanabe}
The top and stop masses are taken
as $m_t$ =~170 GeV and $m_{stop}$ =~1 TeV.
In this figure the Higgs mass is given as a function of $m_A$.
For comparison, the SM (dashed) and the NT
(dotted) cross-sections with the same Higgs mass are also shown.
\label{fig:fig9}}
\end{figure}
Here the top-quark and stop-squark mass are taken as 170 GeV and 1TeV
respectively. In this figure the
lightest Higgs mass varies as a function of $m_A$ (see Figure 3).
For comparison
the production cross-sections for the SM and for a model without
a triple Higgs coupling (NT) calculated with the same Higgs mass
are shown.  We can see that, in the region
of small $m_A$, the cross-section becomes very large compared
to the SM value. This corresponds to a region of parameter space
where $e^+e^- \rightarrow HZ$
and $H \rightarrow hh$ are possible.  On the other hand, there are some
regions of parameter space where the production
cross-section is much reduced compared to the SM.

When the heavy Higgs boson is directly produced and multi-Higgs
production becomes large, an interesting possibility arises
for measuring the $H$-$h$-$h$ coupling constant.\cite{DHZ}
If $2m_h < m_H < 2m_t$, the heavy Higgs boson $H$ can have a
sizable decay branching-ratio both in the $H \rightarrow hh$
and $H \rightarrow WW$ modes.
In such a case the $H$-$h$-$h$ coupling constant can be
extracted from the ratio of
two branching ratios since the coupling of the $H$ to two gauge
bosons is determined from the Higgs production cross section
of the fusion process as well as the Higgs-bremsstrahlung process.

\section{Conclusions}
I have reviewed some aspects of the Higgs physics at future
$e^+ e^-$ linear colliders whose CM energy is ranging from
300 GeV to 1.5 TeV. At earlier stage of the experiment with
$\sqrt{s}\sim$ 300 - 500 GeV, it is easy to find a light
Higgs boson predicted in SUSY standard models or GUT.
In particular, both in the MSSM and the SUSY SM with
a gauge singlet Higgs, at least one of neutral Higgs bosons
is detectable. This is important because it is known that
there is a parameter space in the MSSM
where no signal of Higgs bosons is obtained in the LHC
experiment .

Advantage of linear collider experiments is, however, not only the
discovery potential of the Higgs particle. More importantly,
detailed study on properties of the Higgs boson is possible
through measurements of various production cross-sections and branching
ratios. Here several examples are discussed: anomalous coupling
of $Z$-$Z$-$h$, Higgs couplings to $c\bar{c}$/$gg$/$b\bar{b}$ in
the MSSM, Higss decays to superparticles and the measurement
of triple-Higgs-couplings through the
$e^+e^-$ $\rightarrow Zhh$ process. Combining information obtained
from the LHC experiment, we will be able to clarify the Higgs sector
of the SM and explore physics beyond the SM.

\section*{Acknowledgements}
The author would like to thank A. Djouadi for many useful
discussions, J. Kamoshita and I. Watanabe for helping
to prepare figures, and R. Szalapski for reading the manuscript.

%\newpage

\end{document}